\documentclass[12pt]{iopart}
\usepackage{graphicx,subfigure}
\usepackage[colorlinks=true,linkcolor=blue,citecolor=blue]{hyperref}
\usepackage{color}

\begin{document}

\title{Entropy--energy inequalities for qudit states}

\author{Armando Figueroa$^1$, Julio L\'opez$^1$, Octavio Casta\~nos$^1$,
\\ Ram\'on L\'opez-Pe\~na$^1$ Margarita A. Man'ko$^{2,3}$, Vladimir I. Man'ko$^{2,3}$}

\address{$^1$Instituto de Ciencias Nucleares, Universidad Nacional
Aut\'onoma de M\'exico, Apdo. Postal 70-543 M\'exico 04510 D.F. \\
$^2$ P. N. Lebedev Physical Institute, Leninskii Prospect, 53,
Moscow 119991, Russia \\
$^3$ Moscow Institute of Physics and Technology (State University)
   Dolgoprudnyi, Moscow Region 141700, Russia}

\ead{ocasta@nucleares.unam.mx}

\begin{abstract}
We establish a procedure to find the extremal density matrices for
any finite Hamiltonian of a qudit system. These extremal density
matrices provide an approximate description of the energy spectra of
the Hamiltonian. In the case of restricting the extremal density
matrices by pure states, we show that the energy spectra of the
Hamiltonian is recovered for $d=2$ and $3$.  We conjecture that by
means of this approach the energy spectra can be recovered for the
Hamiltonian of an arbitrary finite qudit system. For a given qudit
system Hamiltonian, we find new inequalities connecting the mean
value of the Hamiltonian and the entropy of an arbitrary state. We
demonstrate that these inequalities take place for both the
considered extremal density matrices and generic ones.
\end{abstract}



\section{Introduction}

Recently~\cite{moreno,cas3,cas4,cordero} an approach was established
to study the ground state properties of algebraic Hamiltonians. This approach follows closely the algorithm established in~\cite{gilmore1,gilmore2}. In
particular, the approach was applied to describe the ground state of
even--even nuclei within the interacting boson model~\cite{moreno}.
In quantum optics, the procedure was used to determine the phase
diagrams of the transitions between the normal regime to the
super-radiant behavior of the ground states of two- and three-level
systems interacting with a one-mode radiation
field~\cite{cas3,cas4,cordero}.  This approach evaluates the mean
value of the Hamiltonian with respect to variational test coherent
states associated to the corresponding algebraic structures of the
Hamiltonian. There exists a tomographic approach, which also uses
mean values of density operators in an ensemble of bases to get
information on the state of the system~\cite{bertrand,vogel,mancini,
ibort, cas5}. For continuous variables, the tomographic approach has
been introduced in~\cite{bertrand,vogel} in the form of optical
tomography. The symplectic tomography is established
in~\cite{mancini}, and a recent review of these tomograms is given
in~\cite{ibort}. The discrete spin tomography has been introduced
in~\cite{dodonov,olga}, while the kernel for product of spin
tomograms is presented in~\cite{lopez1,lopez2}.  The squeezed
tomography is discussed in~\cite{cas5}, which is a fair probability
distribution of a discrete random variable.

One of the aims of this work is to extend the approach mentioned
above to have information of the complete energy spectrum by
considering the mean values of the Hamiltonian with respect to
extremal density matrices~\cite{landau,neumann}. This is achieved by
writing the mean value of the Hamiltonian as a function of the
variables of a general finite-dimensional density
matrix~\cite{tilma, kimura, byrd, jarls, bruning, aktha} together
with the parameters of the Hamiltonian. To guarantee the positivity
of the density matrix, we need to include $d-1$ parameters related
to the purity of the density matrix~\cite{kimura,bruning}.

Another goal of this work is to obtain new inequalities connecting
entropy and mean value of energy for this qudit system.  We show
that there exists a bound for the sum of energy and entropy
determined by the partition function taken for a particular value of
its argument. The method to obtain these inequalities is based on
known property of positivity of the relative entropy involving two
density matrices of the system states~\cite{nielsen}.  Analogous
mathematical inequalities have been discussed
in~\cite{chernega,manko}. The results obtained are valid  for
generic quantum states (qudits).

The main contribution of our work is to demonstrate the new approach
related to the determination of the extremal points of mean values
of the Hamiltonian by considering a general parametrization of the
density matrices for qudit systems and to test the new
entropy--energy inequality.  This inequality contains the bound
determined by the partition function~\cite{kubo}. The formulated
results can be generalized to study the relations between the
entropy and an arbitrary hermitian operator describing an
observable.

\section{Unitary parametrization  of the Hamiltonian operator}

The Hamiltonian operator $\hat{H}$ can be expanded in terms of the
set of operators $\{\hat{\lambda}_1\ldots\hat{\lambda}_{d^2-1}\}$
that form a basis of ${\rm SU}(d)$ and the identity operator
$\hat{I}$ as follows~\cite{kimura}:
\begin{equation}\label{eq:eq1}
\hat{H}=\frac{1}{d} h_0\widehat{I}+ \frac{1}{2} \sum_{k=1}^{d^2-1}
h_k\hat{\lambda}_k \, ,
\end{equation}
with the definitions $h_0 \equiv {\rm Tr}(\hat{H})$ and  $h_k \equiv
{\rm Tr} (\hat{H}\hat{\lambda}_k)$. The generators of ${\rm SU}(d)$
satisfy the relations
\begin{equation}
\hat{\lambda}_k = \hat{\lambda}^\dagger_k \, , \quad {\rm Tr}
( \hat{\lambda}_k) =0 \, , \ {\rm and} \quad {\rm Tr} (\hat{\lambda}_k \,
\hat{\lambda}_j) = 2 \, \delta_{k j} \, .
\end{equation}
They are completely characterized by means of the commutation and
anticommutation relations given in terms of the symmetric and
antisymmetric structure constants of the special unitary group in
$d$ dimensions~\cite{kimura}.

In a similar form, the density matrix can be expanded, i.e.,
\begin{equation}\label{eq:eq3}
\hat{\rho}=\frac{1}{d}\hat{I}+ \frac{1}{2}\sum_{k=1}^{d^2-1}\lambda_k
\, \hat{\lambda}_k \, ,
\end{equation}
because Tr$(\hat{\rho})=1$, and in this case one defines
\begin{equation}\label{eq:eq4}
\lambda_k \equiv {\rm Tr} (\hat{\rho}\hat{\lambda}_k) \, .
\end{equation}

Our purpose is to find the extreme values for the
$\{\lambda_1,\ldots, \lambda_{d^2-1} \}$ variables of the density
matrix by taking the expectation value of the Hamiltonian operator.
To guarantee the positivity of the density matrix, it is necessary
to introduce $d-1$ parameters. Therefore, the extremes are obtained
by means of the definition of a new function depending on
$\lambda_k$ variables with $k=1,2,\dots d^2-1$, $\Lambda_j$ Lagrange
multipliers with $j =2, \dots d$, $h_i$ parameters of the
Hamiltonian with $i=0,
\dots d^2-1$, and $c_l$ real constants  with $l=2 \dots d$
characterizing the  purity of the density matrix
\begin{equation}\label{eq:eq5}
f(\lambda_k,  \Lambda_j, h_i, c_l ) \equiv {\rm Tr}(\hat{H} \,
\hat{\rho}) + \sum_{j=2}^d \Lambda_j (c_j - a_j ) \, ,
\end{equation}
where $a_j$ are nonholonomic constrictions from the characteristic
polynomial of $\hat{\rho}$, which can be obtained by means of the
recursive relation~\cite{bruning}
\begin{equation}\label{eq:eq6}
a_j = \frac{1}{j} \Bigl((-1)^{j-1} {\rm Tr} \, (\hat{\rho}^j) +
\sum^{j-1}_{n=1} (-1)^{n-1} \, a_{j-n} {\rm Tr} (\hat{\rho}^n) \Bigr) \, ,
\end{equation}
where $a_0 \equiv 1$, $a_1 = {\rm Tr} (\hat{\rho}) = 1$, and $j=2,
\dots d$. The parameters $c_j \geq 0$ are constants. To find the
extrema, we derive the function $f(\lambda_k,  \Lambda_j, h_i, c_l
)$  with respect  to $\lambda_q$ obtaining $d^2-1$ algebraic
equations regarding the independent variables of the density matrix.
Then by substituting expressions (\ref{eq:eq1}) and (\ref{eq:eq3})
into (\ref{eq:eq5}), one arrives at
\begin{equation}\label{eq:eq7}
\frac{\partial}{\partial \lambda_q} \, \Biggl(\frac{h_0}{d} + \frac{1}{2}
\sum^{d^2-1}_{k=1} h_k \, \lambda_k \Biggr)   - \sum^d_{j=2} \Lambda_j
\frac{\partial a_j}{\partial \lambda_q}=0  \, ,
\end{equation}
plus $d-1$ differential equations regarding Lagrange multipliers
\begin{equation}\label{eq:eq8}
\frac{\partial}{\partial \Lambda_p}\, f(\lambda_k,  \Lambda_j, h_i, c_l )  =0  \Rightarrow c_p = a_p \, ,
\end{equation}
with $q=1, \dots d^2-1$, $p=2, \dots d$, and we have used the
properties of the generators $\hat{\lambda}_k$ of the unitary group
in $d$ dimensions. These sets of algebraic equations determine the
extremal values of the density matrix, i.e., $\lambda_q=\lambda^c_q$
and $\Lambda_q=\Lambda^c_q$ for which the expressions (\ref{eq:eq7})
and (\ref{eq:eq8}) are satisfied.

\section{Extremal density matrices for $d=2$ and $3$}

\subsection{ Case $d=2$}

One has three generators $\hat{\lambda}_k$ with $k=1, 2, 3$, which
can be realized in terms of the Pauli matrices. Therefore, the
density matrix can be written in the form
\begin{equation} \label{eq:eq9a}
 \hat{\rho}=\frac{1}{2} \left(
\begin{array}{cc}
    1+\lambda_3 & \lambda_1-i\,\lambda_2 \\
    \lambda_1+i\,\lambda_2 & 1-\lambda_3
\end{array} \right) \, ,
\end{equation}
and similarly an arbitrary $2 \times 2$ Hamiltonian matrix is given by
\begin{equation} \label{eq:eq9b}
 \hat{H}=\frac{1}{2} \left(
\begin{array}{cc}
    h_0+h_3 & h_1-i\,h_2 \\
    h_1+i\,h_2 & h_0-h_3
\end{array} \right) \, .
\end{equation}
Substituting the last expressions into Eqs.~(\ref{eq:eq5}), we
obtain
\begin{eqnarray}\label{eq:eq9}
\fl
 f(\lambda_k,\Lambda , h_j, c_2 )=
\frac{1}{2} ( h_0 + h_1 \lambda_1 + h_2 \lambda_2 + h_3 \lambda_3 )
 +  \Lambda (c_2 - \frac{1}{4} (1 - \lambda_1^2 - \lambda_2^2 - \lambda_3^2) ) \, ,
\end{eqnarray}
yielding, by means of expressions (\ref{eq:eq7}) and (\ref{eq:eq8}),
the system of equations
\begin{eqnarray}\label{eq:eq10}
h_k + \Lambda \lambda_k = 0  \, , \quad  \frac{1}{4} \Biggl(1 - \lambda_1^2 -
\lambda_2^2 - \lambda_3^2\Biggr) =c_2   \, ,
\end{eqnarray}
with $k=1,2$ and $3$.  Solving this system of equations, one obtains the results
\begin{equation}\label{eq:eq11}
\lambda^c_k =\mp \frac{\delta}{h}\, h_k \, , \quad \Lambda^c = \pm \, \frac{h}{\delta} \, ,
\end{equation}
with $k=1,2, 3$ and we defined the parameters $h =\sqrt{
h_1^2+h_2^2+h_3^2}$ and $\delta=\sqrt{1-4c_2}$. Therefore, we have
two solutions and substituting them into the expression for the
density matrix, we obtain
\begin{equation}\label{eq:eq15}
\hat{\rho}^c_{\pm} =  \frac{1}{2} \left(
\begin{array}{cc}
 1 \mp {\delta \, h_3}/{h}  & \mp {\delta \, (h_1 -  i h_2 )}/{ h} \\
 \mp {\delta \, (h_1 + i h_2 )}/{ h} &  1\pm {\delta \,  h_3}/{h} \\
\end{array}
\right)  \, .
\end{equation}
Therefore, the extremal density matrices depend on the parameter
$c_2$ whose value is bounded, $0\leq c_2\leq1/4$ and, if the density
matrix represents a pure or a mixed state, it is determined. For
$c_2=1/4$, one has that $\delta=0$ and the extremal density matrix
takes the form
\begin{equation}\label{eq:eq18}
\hat{\rho}^c_{\pm}  = \left(
\begin{array}{cc}
 {1}/{2} & 0 \\
 0 & {1}/{2}
\end{array}
\right)
\end{equation}
corresponding to a mixed state with maximum entropy with the
expectation value of the Hamiltonian given by $\langle H
\rangle_c=h_0/2$.

For $c_2=0$, one gets $\delta=1$ and the expectation values of the
Hamiltonian are given by
\begin{eqnarray}
\langle H \rangle^c_\pm  = \frac{1}{2} \left(h_0 \pm h \right) \, ,
\label{eq:eq19}
\end{eqnarray}
which corresponds exactly to the eigenvalues of the arbitrary matrix
Hamiltonian in (\ref{eq:eq9b}). The corresponding pure states are
given by the density matrices in Eq.~(\ref{eq:eq15}) replacing the
value of the parameter $\delta=1$. They are orthogonal projectors as
it can be proved by multiplying the corresponding density matrices
and taking the trace operation. Therefore, we have reconstructed the
Hamiltonian matrix by finding the extremal values of the expectation
value of the Hamiltonian.

\begin{figure}[ht]
\begin{center}
    \includegraphics[scale=1]{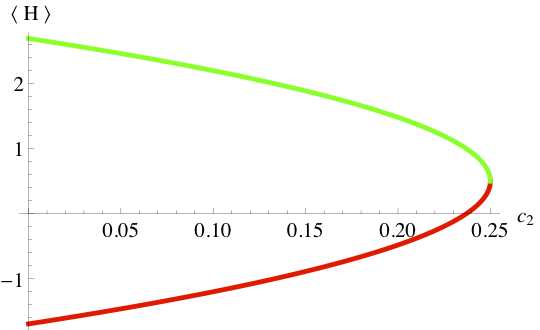}
\caption{\label{fig1}
Expectation value of the two-level-atom Hamiltonian as a function of
$c_2$, with $h_0=1$, $h_1 = \sqrt 2$, $h_2 = e$, and $h_3 = \pi$.}
\end{center}
\end{figure}

In Fig.~\ref{fig1}, we plot the general behavior of the expectation
value of a given Hamiltonian operator $\hat{H}$ as a function of the
parameter $c_2$. Notice that if $c_2 \approx 0$, the mean values of
the energy with respect to mixed states are close to the maximum and
minimum values of the energy spectra. Additionally, we can observe
that for each value of $c_2$ there are two solutions for the
expectation values of the Hamiltonian, except when $c_2=1/4$, where
one gets only one expectation value associated to the mixed state
with maximum entropy.
This procedure gives the eigenvalues and eigenvectors of the
Hamiltonian and besides one gets information on the mean values
associated to mixed states, as function of $c_2$.

\subsection{Case $d=3$}

We consider a particular Hamiltonian describing a two-mode
Bose--Einstein condensate~\cite{milburn,citlali},
\begin{equation}\label{eq:eq23}
\hat{H}= a\, \hat{J}_z + b \hat{J}_{z}^2 + c \hat{J}_x,
\end{equation}
where $\hat{J}_k$ denotes the $k$-th component of the angular
momentum operator. The parameter $a$ corresponds to the difference
in the chemical potentials between the wells, $b$ represents the
atom--atom interaction, and $c$ is related to the atom tunneling
parameter.

For the qutrit case, one substitutes the three-dimensional
representation of the angular momentum operators, and the generators
$\hat{\lambda}_{k}$, with $k=1,2, \cdots, 8$ can be realized in
terms of the Gell-Mann matrices. Thus, an arbitrary density matrix
is denoted by
\begin{equation}
    \hat\rho =\left(
    \begin{array}{ccc}
    \frac{1}{3}+\frac{1}{2}\left(\lambda_{7}+\frac{1}{\sqrt{3}}\lambda_{8}\right)&
    \frac{1}{2}\left(\lambda_{1}-i\,\lambda_{4}\right)&
    \frac{1}{2}\left(\lambda_{2}-i\,\lambda_{5}\right)\\
    \frac{1}{2}\left(\lambda_{1}+i\,\lambda_{4}\right)&
    \frac{1}{3}-\frac{1}{2}\left(\lambda_{7}-\frac{1}{\sqrt{3}}\lambda_{8}\right)&
    \frac{1}{2}\left(\lambda_{3}-i\,\lambda_{6}\right)\\
    \frac{1}{2}\left(\lambda_{2}+i\,\lambda_{5}\right)&
    \frac{1}{2}\left(\lambda_{3}+i\,\lambda_{6}\right)&
    \frac{1}{3}-\frac{1}{\sqrt{3}}\lambda_{8}\
    \end{array}
    \right) \ ,
\end{equation}
and one has a similar expression for the Hamiltonian. Comparing with
the $j=1$ angular momentum representation, one obtains that the
parameters different from zero are
\begin{equation}
h_{0}=2\,b,\quad h_{1}=h_{3}=\sqrt{2}\,c,\quad h_{7}=a+b,\quad
h_{8}=({3\,a-b})/{\sqrt{3}}\ .
\end{equation}
Substituting the previous expressions for the density matrix and the
Hamiltonian into Eq.~(\ref{eq:eq5}), we obtain the function $f$ in
the form
\begin{eqnarray}\label{eq:eq24}
&&
\fl
f(\lambda_k, \Lambda_2,\Lambda_3,a,b,c,c_2,c_3) =
 \nonumber\\
&&
\fl
\qquad\frac{2}{3}b+\frac{\sqrt{2}}{2}c\left(\lambda_1+\lambda_3\right)+
\frac{1}{2}\left(a+b\right)\lambda_7+\frac{\sqrt{3}}{6}\left(3a-b\right)
\lambda_8+{\Lambda_2} \biggl(c_2 -\frac{1}{3}  + \frac{1}{4} \,\lambda^2
\biggr) \nonumber\\
&&
\fl
\quad+\Lambda_3\biggl(c_3- \frac{1}{27}+\frac{1}{12}\lambda^2 -\frac{1}{4}
\left(\lambda_1\lambda_2\lambda_3+\lambda_3\lambda_4\lambda_5-\lambda_2
\lambda_4\lambda_6+\lambda_1\lambda_5\lambda_6\right)+\frac{1}{12\sqrt{3}}
\lambda_8^3 \nonumber\\
&&
\fl
\quad-\frac{1}{8}\left(\lambda_2^2-\lambda_3^2+\lambda_5^2-\lambda_6^2\right)
\lambda_7 -\frac{1}{8\sqrt{3}}\left(2\lambda_1^2-\lambda_2^2-\lambda_3^2+
2\lambda_4^2-\lambda_5^2-\lambda_6^2+2\lambda_7^2\right)\lambda_8 \biggr)\ ,
\end{eqnarray}
where we define
$\lambda=\sqrt{\lambda^2_1+\lambda^2_2+\lambda^2_3+\lambda^2_4
+\lambda^2_5+\lambda^2_6+ \lambda^2_7 + \lambda^2_8}$. The extrema
of the previous function with respect to
$\lambda_1,\ldots\lambda_8,\Lambda_2,\Lambda_3$, are obtained
numerically by establishing different values of the set of
parameters $\{a, b, c, c_2, c_3\}$.  Notice that the $c_2$ and $c_3$
are not independent parameters, and they are keeping track of the
purity of the extremal density matrices.  If we define $\tau_2 =
\tr{\rho^2}$ and $\tau_3=\tr{\rho^3}$, then following~\cite{gerdt},
it is straightforward to get the compatible region of the parameters
$\tau_2$ and $\tau_3$. For the set $(c_2, c_3) = (1/3, 1/27)$, one
gets the extremal density matrix, $\rho^c=\hat{I}/3 $, for the mixed
state with maximum entropy and $\langle H \rangle_c = 2\, b /3$.

\begin{figure}[ht]
     \begin{center}
        \subfigure[$\langle \hat{H} \rangle$ as a function of the parameter $a$, with $b~=~0.5$, and $c~=~-1$.]{%
            \label{fig2a}
            \includegraphics[width=0.45\textwidth]{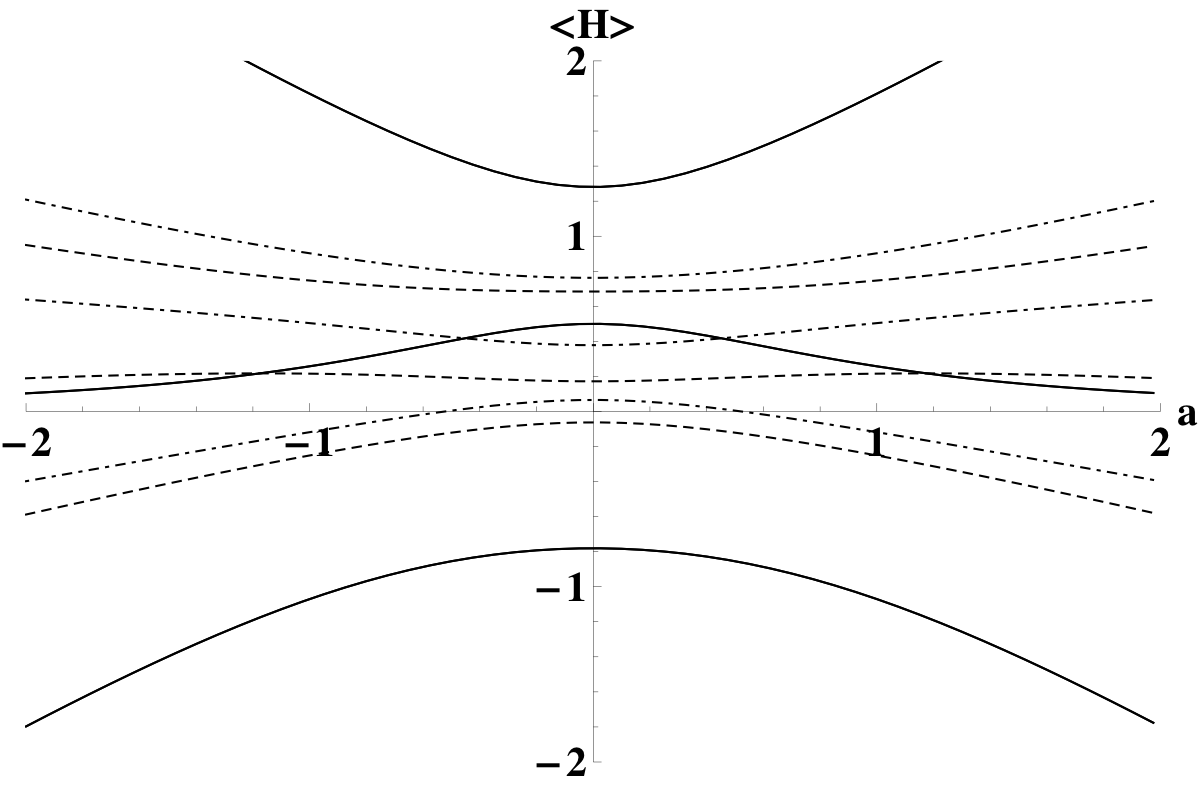}
        } \quad
        \subfigure[$\langle \hat{H} \rangle$ as a function of the parameter $b$, with $a=0.5$, and $c=0.5$.]{%
           \label{fig2b}
           \includegraphics[width=0.45\textwidth]{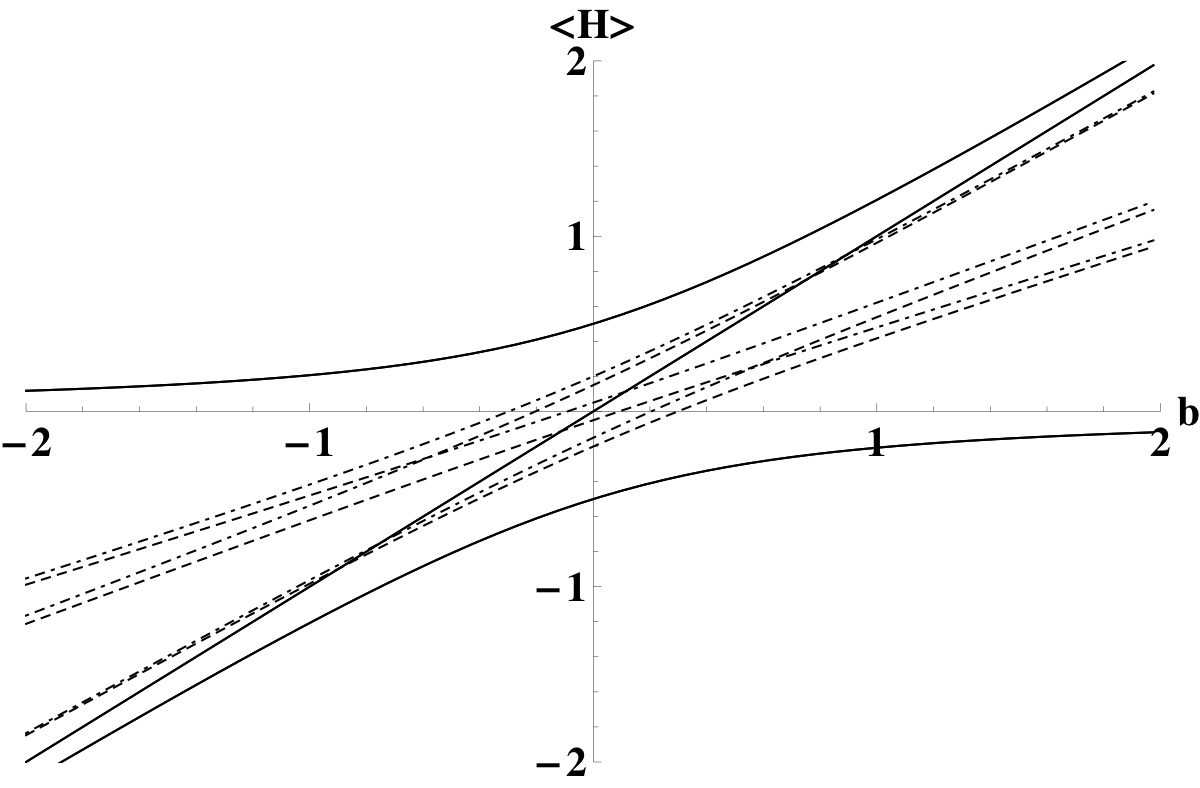}
        }\\ 
        \subfigure[$\langle \hat{H} \rangle$ as function of $c$, with $a=0.5$, $b=-1$.]{%
            \label{fig2c}
            \includegraphics[width=0.45\textwidth]{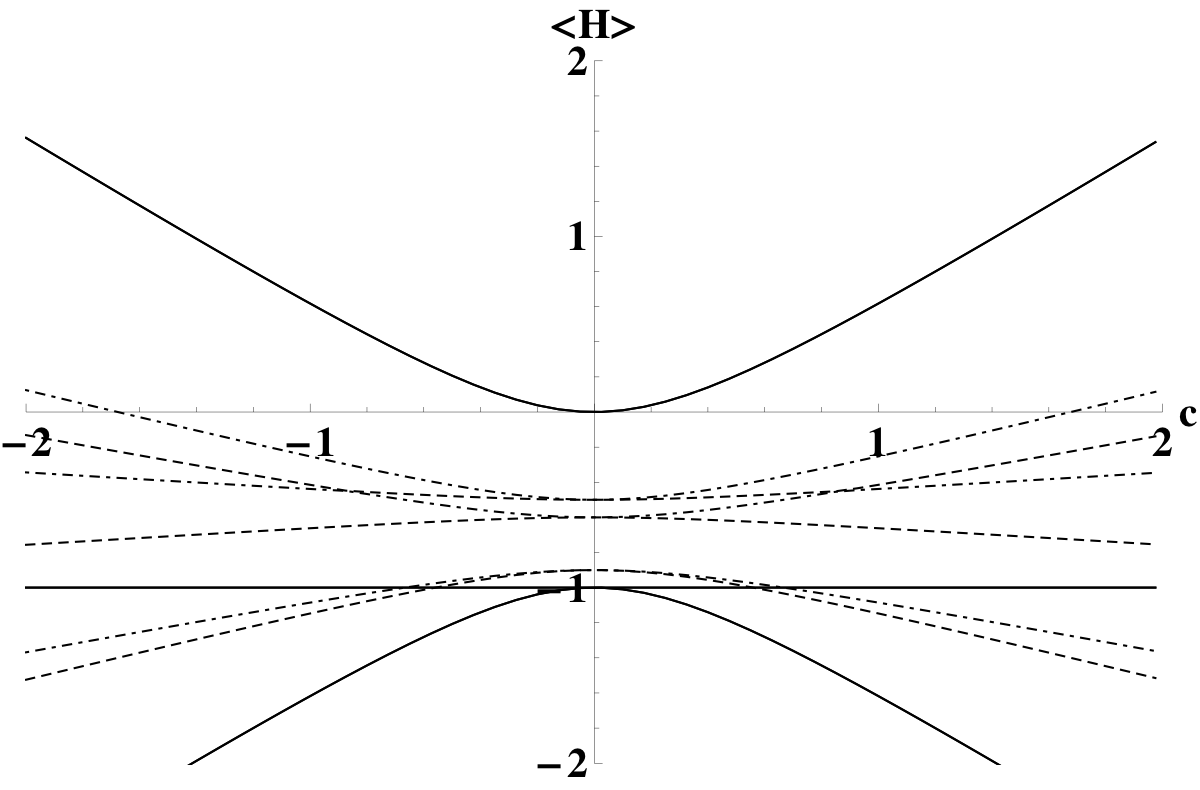}
        } \quad
        \subfigure[$\langle \hat{H} \rangle$ as a function of $a$, with $b=-0.5$, $c=-1$.]{
            \label{fig2c}
            \includegraphics[width=0.45\textwidth]{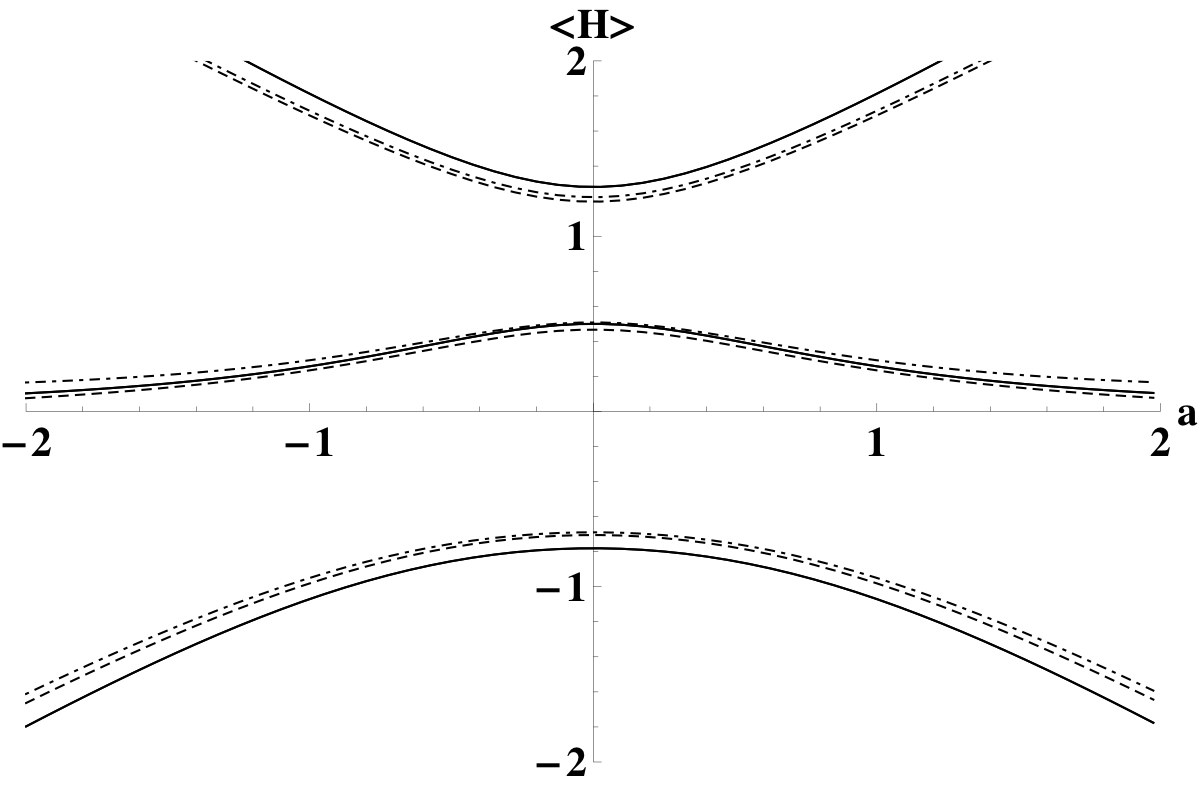}
        }
    \end{center}
    \caption{%
Three different plots of the expectation value $\langle \hat{H}
\rangle$ as a function of the parameters of the Hamiltonian
for $(c_2= \frac{29}{100}$ and $c_3=\frac{1}{50})$~(a--c) and $(c_2=
\frac{1921}{40000}$ and $c_3=\frac{399}{800000})$~(d) in the domain
determined in~\cite{gerdt}. They are indicated by dashed lines and
dash-dotted lines; notice that there are two independent solutions,
each of one with three curves.  The three black curves correspond to
the pure states of the system, where $c_2= c_3=0$.
     }%
   \label{fig2}
\end{figure}

To illustrate the procedure, we consider three different choices of
the mentioned set of parameters. The obtained expectation values of
the Hamiltonian are plotted in Fig.~\ref{fig2} as functions of $a$,
$b$, $c$.  The pure states are associated to $c_2=c_3=0$, and we
find that there are $3$ extremal solutions for the parameters of the
density matrix which are indicated by black continuous lines.
Comparing these different mean values of the Hamiltonian with the
corresponding exact diagonalizations, one finds a complete agreement
with the energy spectra.

For $c_2$ and $c_3$ different from zero, additionally one gets
information on the mean values of the energy.  We obtain in this
case that the number of mean values of mixed states is related to
the symmetric group of $3$ dimensions; thus, for $c_2=
\frac{29}{100}, c_3=\frac{1}{50}$, there are $6$ colored continuous
lines indicating the expectation values of the Hamiltonian of the
possible mixed states.

\section{Entropy and energy relations}

It is known~\cite{nielsen} that for two given density matrices
$\rho$ and $\sigma$, the positivity condition for their relative
entropy is given by
\begin{equation} \label{eq:eq25}
    {\rm Tr}(\rho \, {\ln} \rho - \rho \ln \sigma) \geq 0 .
\end{equation}
The matrices $\rho$ and $\sigma$ satisfy the properties
$\rho^{\dagger} = \rho$, $\sigma^{\dagger} = \sigma$, $\rho \geq 0$,
$\sigma \geq 0$ and ${\rm Tr}\,\rho={\rm Tr} \, \sigma=1$. On the other hand, two arbitrary matrices which
have these properties satisfy inequality (\ref{eq:eq25}).

We are going to use this inequality considering the Hamiltonian
matrix $H$ of a qudit system, by defining the matrix
\begin{equation} \label{eq:eq26}
    \rho_{H} = \frac{e^H}{{\rm Tr} \, e^H} ,
\end{equation}
which has the properties Tr $\rho_{H} = 1$ and $\rho_{H} \geq 0$.

Let us write the positivity conditions of relative entropy of two
matrices $\rho$ and $\rho_{H}$
\begin{equation} \label{eq:eq28}
    {\rm Tr} \left( \rho \ln \rho - \rho \ln \frac{e^H}{{\rm Tr}e^H}
    \right) \geq 0 .
\end{equation}
Using the definition of the von Neumann entropy $S$ associated with
the density matrix of the state, which reads $S = -  {\rm Tr}(\rho \ln \rho)$,
one can rewrite inequality (\ref{eq:eq28}) in the form
\begin{equation} \label{eq:eq30}
    S + \langle H \rangle \leq \ln \left ({\rm Tr} e^H \right) ,
\end{equation}
where the mean energy is given by $E=\langle H \rangle$.

For an arbitrary Hamiltonian $H$, one introduces the partition
function
\begin{equation} \label{eq:eq32}
    Z(\beta) = {\rm Tr} \left(e^{- \beta H} \right),
\end{equation}
where $T=\beta^{-1}$ is interpreted as a temperature, and the matrix
$\rho(\beta) = \exp(- \beta H)/Z(\beta)$ can be interpreted as the
density matrix of the system in thermal equilibrium state. It is
known that the partition function $Z(\beta)$ determines all
thermodynamic properties of the system (cf.~\cite{kubo}). Then
inequality (\ref{eq:eq30}) can be rewritten in the form
\begin{equation} \label{eq:eq33}
    E + S \leq \ln Z(\beta=-1) .
\end{equation}
Thus, we got for an arbitrary quantum system a bound for the sum of
the mean energy value and the von Neumann entropy, and this bound is
determined by the partition function evaluated in a particular value
of temperature, $T=-1$.

For example, for a two-level atom in the state determined by the
density matrix given in Eq.~(\ref{eq:eq9a}) with the condition
$\lambda^2_1 + \lambda^2_2 + \lambda^2_3 \leq 1$, and the
Hamiltonian matrix given in Eq.~(\ref{eq:eq9b}), inequality
(\ref{eq:eq33}) reads
\begin{eqnarray}
\label{eq:eq36}
&& \frac{1}{4} \, {\rm Tr} \left\{ \left(
\begin{array}{cc}
    1+\lambda_3 & \lambda_1-i\,\lambda_2 \\
    \lambda_1+i\,\lambda_2 & 1-\lambda_3
\end{array}
\right)
\left(
 \begin{array}{cc}
    h_0+h_3 & h_1-i\,h_2 \\
    h_1+i\,h_2 & h_0-h_3
\end{array}
\right) \right\}  \nonumber \\
&& -  {\rm Tr} \left\{ \left(
\begin{array}{cc}
     \frac{1+\lambda_3}{2} & \frac{\lambda_1-i\,\lambda_2}{2} \\
    \frac{\lambda_1+i\, \lambda_2}{2} & \frac{1-\lambda_3}{2}
\end{array}
\right) \ln
\left(
\begin{array}{cc}
   \frac{1+\lambda_3}{2} & \frac{\lambda_1-i\,\lambda_2}{2} \\
    \frac{\lambda_1+i\, \lambda_2}{2} & \frac{1-\lambda_3}{2}
\end{array}
\right) \right\} \leq \ln \, {\rm Tr} \, (e^H) \, .
\end{eqnarray}
Notice that the last expression is invariant under unitary
transformations and thus one can write ${\rm Tr} (e^H) = e^{E_{1}} +
e^{E_{2}}$, where the energy levels $E_{1}$ and $E_{2}$ are the
solutions of the secular equation ${\rm Det}( H- E I_{2} ) =0$.

\begin{figure}
\begin{center}
    \includegraphics[scale=0.5]{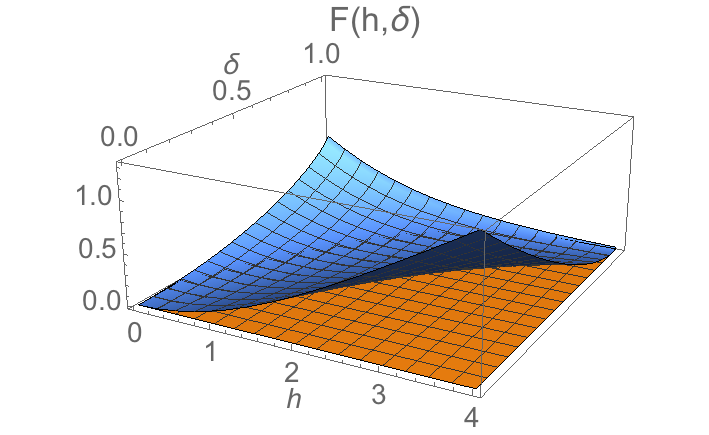} \\[2mm]
    \includegraphics[scale=0.5]{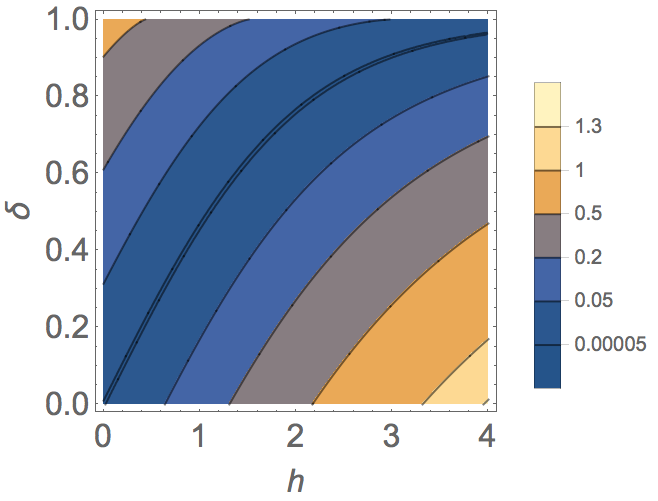}
\caption{\label{fig33}
The entropy--energy inequality~(at the top) and a contour level plot
of this inequality as a function of $h$ and $\delta$~(at the
bottom).}
\end{center}
\end{figure}

For any state, one has that the inequality
(\ref{eq:eq36}) can be written as follows
\begin{equation} \label{eq:eq37}
F=\ln Tr (e^H)- Tr(\rho H) -S \geq 0 ,
\end{equation}
where we define the function $F$ in terms of $(\lambda_1,\lambda_2,\lambda_3)$ and $(h_0,h_1,h_2,h_3)$. By substituting the extremal parameters of the density matrix given
in expression~(\ref{eq:eq11}) the function $F$ takes the form
\begin{equation}
\fl
F(h, \delta) = \frac{1}{2} \left(  -\delta \, h + 2 \, \delta \,
\hbox{arctanh}{(\delta)} + 2 \, \ln {\left(\cosh{\frac{h}{2}}\right)}+
\ln{(1-\delta^2)}\right) \geq 0,
\end{equation}
where $(h,\delta)$ were
defined after Eq.~(\ref{eq:eq11}). This function is
displayed in Fig.~\ref{fig33} as a function of the $\delta$  and
$h$. Notice that always it is larger than zero. The corresponding
contour levels are also positive.

\begin{figure}
\begin{center}
    \includegraphics[scale=0.35]{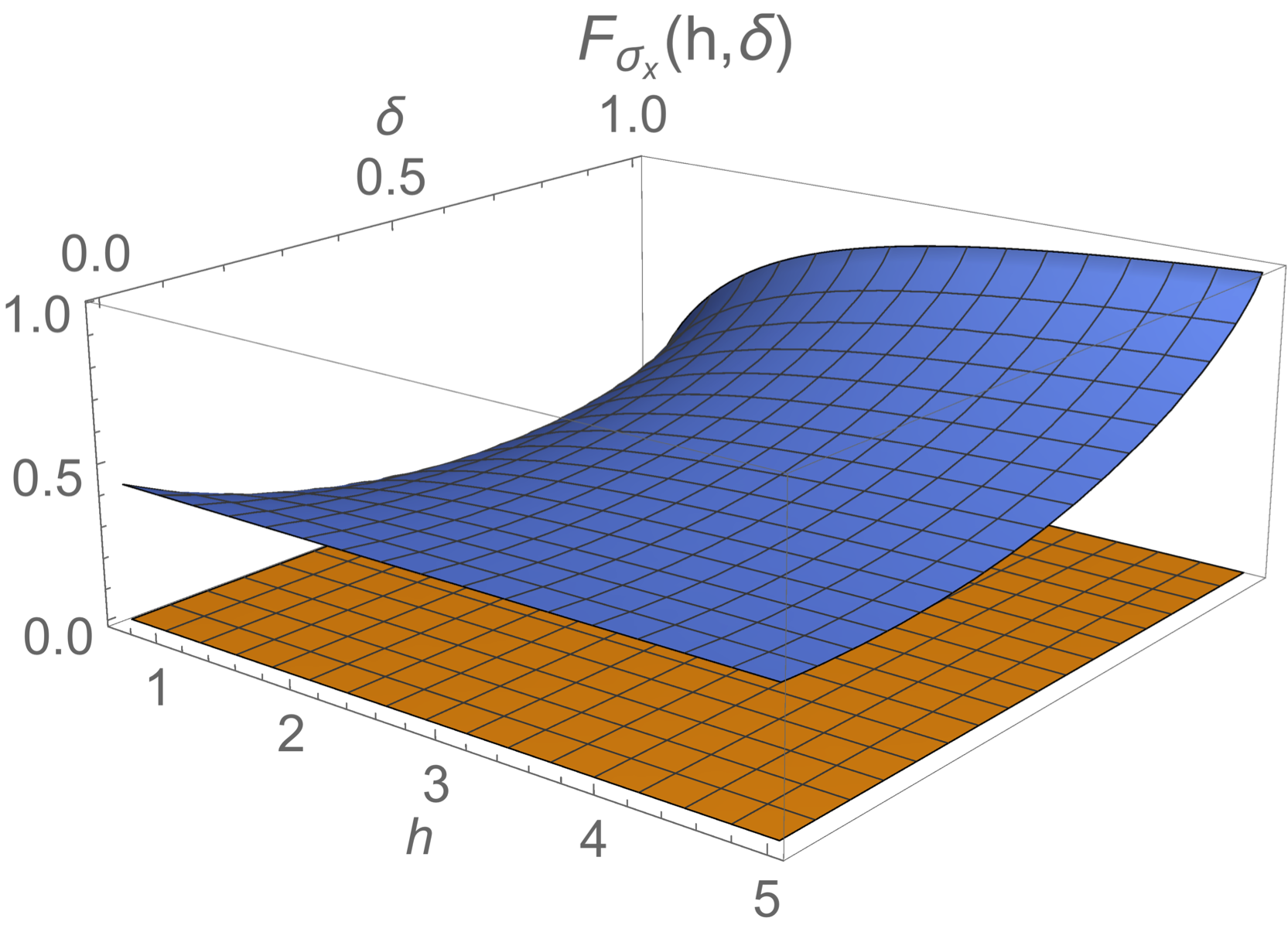} \\[2mm]
    \includegraphics[scale=0.4]{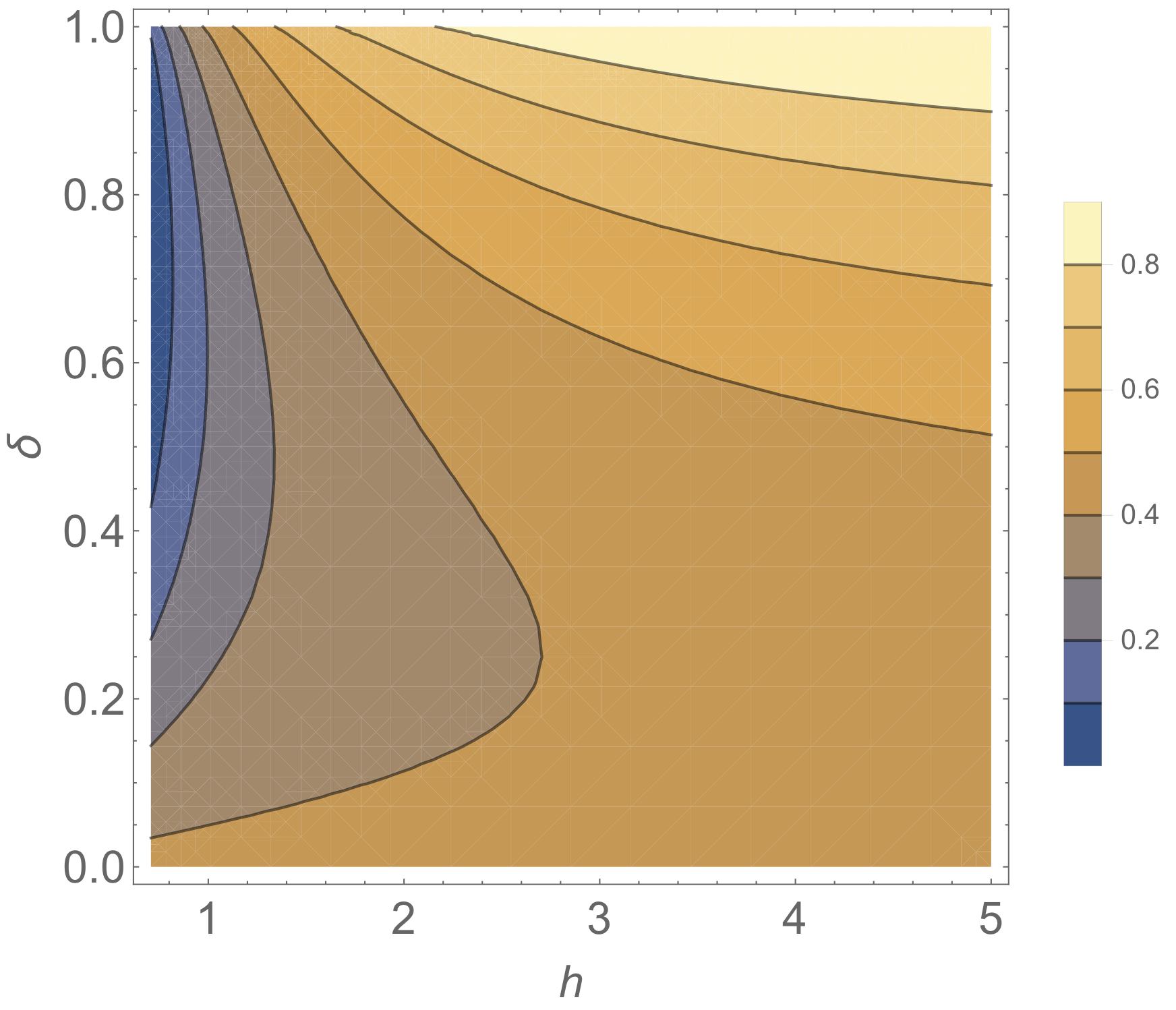}
\caption{\label{fig4}
The entropy and $\sigma_x$ inequality~(at the top) and a contour
level plot of the inequality as a function of $h =\sqrt{\frac{1}{2}
+ h^2_2 + h^2_3}$ and $\delta$~(at the bottom). The expectation
value of $\sigma_x$ is evaluated with respect to one extremal
density matrix of the general Hamiltonian in the two dimensions. }
\end{center}
\end{figure}

Similar inequalities can be obtained for any Hermitian operator, and
the obtained result for the Pauli matrix $\sigma_k$ with $k=x, y,
z$, can be written in the form
\begin{equation}
F_{\sigma_{k}} (h, \delta) = -\frac{ h_k \, \delta}{h} +  \, \delta \,
\hbox{arctanh}{(\delta)}+ \ln{\left(e + 1/e \right)} + \,
\ln{\left(\frac{1-\delta^2}{4}\right)}  \geq 0 \,  ,
\end{equation}
with $k=x,y,z$. For $k=x$, the result is displayed in
Fig.~\ref{fig4}.

One can obtain another important inequality between the energy and the partition function by considering
\begin{equation} \label{eq:eq39}
{\rm Tr} \left\{ \frac{e^H}{{\rm Tr} (e^H)} \ln \left(\frac{e^H}{{\rm Tr}
(e^H)}\right) - \frac{e^H}{{\rm Tr} (e^H)} \ln \rho \right\} \geq 0 ,
\end{equation}
which can be written for any state $\rho$ in the form
\begin{equation} \label{eq:eq40}
- \frac{\partial Z(\beta)}{\partial \beta}\Big\vert_{\beta=-1} -\,
Z(\beta=-1) \, \ln Z(\beta=-1)  \geq {\rm Tr}(e^H \ln \rho) \, .
\end{equation}

Also inequality~(\ref{eq:eq33}) is accompanied by the inequality
$-E+S\leq \ln Z(\beta=1)$, which means that the partition function
provides the bound for the difference of entropy and energy.

\section{Summary and Conclusions}

We suggested to study the properties of a qudit system considering
the mean values of its Hamiltonian with respect to a generic density
matrix, which includes the special cases of pure states. The  number
of parameters which determine the mean values of the Hamiltonian in
this case is equal to the number of parameters determining the
density matrix. Due to this, the information contained in the mean
values of the Hamiltonian is sufficient to reconstruct the
Hamiltonian, including both its spectrum and its eigenvectors. We
demonstrated how the Hamiltonian spectrum is recovered finding all
the extreme density operators which minimize the mean values of the
Hamiltonian. The suggested approach for finding all the properties
of the Hamiltonian has common features with quantum tomography.  In
the quantum tomography approach, one determines the density operator
by making measures in all the reference frames of the information
contained in its mean value. Therefore, the mean values are
calculated for an ensemble of basis determined by sufficient  number
of parameters associated to unitary matrices transforming the
density operator (or equivalently, changing the basis where the
means are calculated). Thus, we found that the description of the
Hamiltonian by the mean values approach and the description of
density operators by tomographic approach have common features. We
demonstrated that for $d=2$ and 3; the Hamiltonian spectrum is
completely recovered using the mean value approach. We surmise on
the basis of the number of parameters of the test variational density matrix that the complete recovering of the spectrum takes
place for arbitrary dimensions. The previous conjecture is based in the following calculation: For pure states a test density matrix with 4 real parameters can be used and then one can get the exact energy spectrum of the system while with the extremal density matrices associated to the CS one has only approximate results for the energy levels.
The problem of the convenience of
our approach in comparison with standard diagonalisation of the
Hamiltonian has to be clarified. We point out that our approach can
be applied to arbitrary hermitian matrices, i.e., to arbitrary
observables. The known properties of density matrices and von
Neumann entropy were used to formulate and verify a new inequality
for a qudit system between the energy and the entropy. We mapped the
Hamiltonian matrix onto a density-like nonnegative matrix, and
applied the property of nonnegativity of relative entropy connecting
two density matrices. The obtained result for the sum (difference)
of entropies and mean value of the Hamiltonian is that this sum
(difference) is bound. The bound is determined by the partition
function evaluated at a specific value of its argument. We verified
the new inequality on an example of a qubit system. The obtained
result is valid for all the states, including entangled states of
multi-qudit systems. One can get analogous inequalities for other
hermitian operators corresponding to physical observables. We will
study this aspect of the inequalities in future publications.

\section{Acknowledgements}
This work was partially supported by CONACyT-M\'exico and PAPIIT-UNAM (IN110114).

\end{document}